\documentclass{PoS}

\newcommand{\be}{\begin{equation}}
\newcommand{\ee}{\end{equation}}
\newcommand{\bea}{\begin{eqnarray}}
\newcommand{\eea}{\end{eqnarray}}

\title{Glueball masses from ratios of path integrals}

\ShortTitle{Glueball masses from ratios of path integrals}

\author{\speaker{Leonardo Giusti}\\
       Dipartimento di Fisica, Universit\'a di Milano-Bicocca, \\
       Piazza della Scienza 3, I-20126 Milano, Italy\\
       E-mail: \email{leonardo.giusti@unimib.it}}

\author{Michele Della Morte\\
       Institut f\"ur Kernphysik and Helmholtz Institut, University of Mainz,\\
       Johann-Joachim-Becher Weg 45, D-55099 Mainz, Germany\\
       E-mail: \email{morte@kph.uni-mainz.de}}

\abstract{\vspace{-11.5cm}
By generalizing our previous work on the parity symmetry, the
partition function of a Yang--Mills theory is decomposed into a sum
of path integrals each giving the contribution from multiplets of states
with fixed quantum numbers associated to parity, charge
conjugation, translations, rotations and central conjugations.
Ratios of path integrals and correlation functions can then be
computed with a multi-level Monte Carlo integration scheme whose
numerical cost, at a fixed statistical precision and at asymptotically
large times, increases power-like with the time extent of the lattice.
The strategy is implemented for the SU(3) Yang--Mills theory, and a
full-fledged computation of the mass and multiplicity of the lightest
glueball with vacuum quantum numbers is carried out at two values
of the lattice spacing (0.17 and 0.12 fm).
\vspace{-15.9cm}
\begin{flushright}
HIM-2011-09 
\end{flushright}
}

\FullConference{The XXIX International Symposium on Lattice Field Theory - Lattice 2011\\
            July 10-16, 2011\\ 
		Squaw Valley, Lake Tahoe, California}

\begin{document}
\section{Introduction}
\vspace{-0.375cm}

Very often the signal-to-noise ratio of correlation functions computed
by ``standard'' Monte Carlo techniques decreases exponentially with the time
separation of the sources~\cite{Parisi:1983ae,Lepage:1989hd}. In spectrum 
computations one has thereby to find compromises between large statistical 
errors at large time-distances and large systematic errors, due to 
contaminations of excited states, at short time-separations. This 
is not entirely satisfactory from a theoretical point of view since
a solid evidence that a single state dominates, i.e.  a long exponential decay 
over many orders of magnitude, is usually missing. 

The situation is particularly unfavorable in the case of the glueball spectrum, 
as the variance of two-point functions at large time-distance is dominated 
by the vacuum~\cite{Parisi:1983ae,Lepage:1989hd}. An intuitive way to understand 
the problem starts from the observation that symmetries are usually not preserved 
on a single gauge configuration. No matter what quantum numbers are specified at 
the source and sink, all states are allowed to propagate in the time direction and
the expected signal emerges in the gauge average as a result of possibly large 
cancellations.

In a series of papers~\cite{DellaMorte:2007zz,DellaMorte:2008jd,DellaMorte:2010yp} 
(see also \cite{DellaMorte:2009rf,DellaMorte:2010ke}) we have proposed and tested a 
computational strategy in which, for each configuration, only states with specified 
quantum numbers are allowed to propagate in the time direction. In such a setup 
the signal-to-noise problem can be solved by introducing
a hierarchical integration scheme~\cite{Parisi:1983hm,Luscher:2001up}. We have implemented 
this ``Symmetry Constrained Monte Carlo'' for computing the mass and the 
multiplicity of the lightest glueball state in the SU(3) Yang--Mills theory.
Here we briefly review the basic ingredients entering that computation, the results 
at a rather coarse value of the lattice spacing ($a$=0.17 fm), and we present new
numerical data at an additional finer resolution ($a$=0.12 fm).
\vspace{-0.20cm}

\section{Lattice symmetries}
\vspace{-0.375cm}

We aim at computing the ratio of the partition function restricted to a sector, identified by a
complete set of conserved quantum numbers, over the standard one. To this end, and to fix the 
notation, it is useful to list the symmetry groups of the SU(3) Yang-Mills theory on a finite 
periodic lattice of volume $V=T \times L^3$, where $T$ is its time-extent and $L$ its length 
in each spatial directions\footnote{Dimensionful quantities are always expressed in units of $a$
unless explicitly specified.}. We will denote by $\Gamma^\mu({\cal{R}}_i)$ the matrix associated 
to the $i$-th element of the group in the irreducible representation $\mu$.
\begin{itemize}
\item{\bf Parity.} The group is of order 2, and the two irreducible representations of dimension 1 are 
$\Gamma^{(\pm)}({\mathcal{R}}_1)=\pm\Gamma^{(\pm)}({\mathcal{R}}_2)=1$.
\item{\bf Charge Conjugation.} Again this group is of order 2, and the two irreducible representations of dimension 1 are 
$\Gamma^{(\pm)}({\cal{R}}_1)=\pm\Gamma^{(\pm)}({\cal{R}}_2)=1$.
\item{\bf Translations.} The group of translations is a direct product of three Abelian groups, one
for each space direction. Its elements are labeled
by a three dimensional vector of integers ${\bf m} = ({m_1,m_2,m_3})$,
with ${m_i}=0,\dots,L-1$, where each component identifies the elements
of the group in the corresponding direction. 
Since it is Abelian, each element forms its own class
and there are $L^3$ non-equivalent irreducible representations 
of dimension 1 
\be
\Gamma^{(\bf p)}({\cal{R}}_{\bf m}) = e^{i\, {\bf p \cdot m}}\; ,
\ee 
which are labeled
by momentum vectors $\displaystyle {\bf p}=\frac{2\pi}{L}\, [n_1,n_2,n_3]$, with
$n_i=0,\dots,L-1$. 
\item{\bf Rotations.} The octahedral group is of order 24. Its elements are listed in 
Appendix B of~\cite{DellaMorte:2010yp}, where explicit expressions for the $\Gamma$ matrices and 
the table of characters 
can be found for the 5 non-equivalent irreducible representations. Those are two singlets 
${\rm A}_1$ and ${\rm A}_2$, one doublet ${\rm E}$ and two 
triplets ${\rm T}_1$ and ${\rm T}_2$. 
\item{\bf Central Charge Conjugations.} This symmetry is strictly related to the choice of
periodic boundary conditions, and it disappears 
in the infinite volume limit~\cite{'tHooft:1979uj}. The group is a direct product of 
three $Z_3$, one for each spatial direction. It is of order 27, and 
its elements are labeled by a three dimensional vector 
of integers ${\bf \nu}=(\nu_1,\nu_2,\nu_3)$, with $\nu_i=0,1,2$,
where each component labels the elements of the Abelian 
group in the corresponding direction. Since each element forms its own class,
there are $27$ non-equivalent irreducible representations of dimension one
\be
\Gamma^{(\bf e)}({\cal{R}}_{\bf \nu}) = e^{i\, {\bf e \cdot} {\bf \nu}}\; ,
\ee
which are labeled by the electric flux vectors 
$\displaystyle {\bf e}=\frac{2\pi}{3}\, [e_1,e_2,e_3]$, with $e_i=0,1,2$. 
\end{itemize}
\vspace{-0.625cm}

\section{Symmetry constrained Monte Carlo}
\vspace{-0.375cm}

For a discrete group of order $g$, the projector $\hat{\rm P}_\mu$ onto states 
which transform as an irreducible representation $\mu$ can be defined as 
(for unexplained notation see Ref.~\cite{DellaMorte:2010yp})
\be
\hat{\rm P}^\mu ={{n_\mu}\over{g}} \sum_{i=1}^g \chi^{(\mu)*}({\mathcal{R}}_i) 
\hat{\Gamma}({\mathcal{R}}_i) \;,
\label{eq:proj}
\ee
where $n_\mu$ is the dimension of the irreducible representation, $\chi^{(\mu)}({\mathcal{R}}_i)$
is the character of the $i$th group element in that representation, and 
$\hat{\Gamma}({\mathcal{R}}_i)$ is the representation of the element in the Hilbert space. 
The corresponding symmetry-constrained partition function can then be expressed as
\be
Z^{(\mu)}(T) ={\rm Tr} \left\{ \hat{\rm T}^T \hat{\rm P}^\mu \right\}\;,
\ee
where $\hat{\rm T}$ is the transfer matrix among gauge-invariant states.
By inserting (\ref{eq:proj}) in the equation above  
and by choosing the ``coordinate'' basis to express the trace, 
it is clear that $Z^{(\mu)}(T)$ can be written as a linear combination of partition functions 
of $g$ different systems with twisted boundary conditions in the time direction. The latter are chosen 
so that the state at time $T$ is related to the one at time $0$ by a group transformation. As shown 
in detail in Ref.~\cite{DellaMorte:2008jd}, the ratio $Z^{(\mu)}(T)/Z(T)$ can be factorized as a product 
of similar ratios associated to thick time-slices of temporal extension $d$ and fixed boundary 
conditions. Explicitly, we numerically compute
\be
{{Z^{(\mu)}(T)}\over{Z(T)}}={{1}\over{Z(T)}} \int D U\, e^{-S[U]}\, {\rm P}_{m,d}^{(\mu)}[T,0]\;,
\ee
where ${\rm P}_{m,d}^{(\mu)}[T,0]$ is a product of $m=T/d$ factors, 
and depends on the values of the 
spatial links on the boundaries of the thick time-slices only. Its definition is given 
in Refs.~\cite{DellaMorte:2008jd,DellaMorte:2010yp}. Finally, given the locality of the gauge theory, 
such a factorized quantity can be very accurately estimated through a generalization of the 
hierarchical integration scheme proposed in Refs.~\cite{Parisi:1983hm,Luscher:2001up}, 
which in this case removes completely the exponential signal-to-noise problem 
\cite{DellaMorte:2007zz,DellaMorte:2008jd,DellaMorte:2010yp}.  Although we focus here on partition functions, 
the same approach can be applied to the computation of correlation functions. Matrix elements of 
operators among glueball states can again be obtained avoiding the exponential 
signal degradation~\cite{DellaMorte:2010yp}.

To determine the mass of the lightest glueball state with vacuum quantum numbers, we are interested in 
computing the ratio $Z^{({\bf e}=0,\,{\bf p},\,{\cal C}=+)}(T)/Z(T)$, which we will shorten as 
$Z^{({\bf p},+)}(T)/Z(T)$. The projection onto non-zero momentum is needed to get rid of the contribution 
from the vacuum, which would otherwise dominate the variance and cause an exponential 
degradation of the signal from the glueball. By applying blindly the analysis in this 
section, we would need to calculate the thick time-slice 
ratios for each of the $L^3\times 27\times2$ boundary conditions. This would make the 
approach extremely expensive from the computational point of view, and it would give unnecessary 
information if one is interested in the low momentum states only. It is possible, however, to have 
still an exact numerical algorithm by implementing the projectors on 
$p_x$ and ${\cal C}$ exactly, while treating those on $(p_y,p_z)=(0,0)$ and ${\bf e}=0$ stochastically 
by extracting $n$ random transformations out of the associated $L^2\times 27$ ones and averaging 
over them, see Ref.~\cite{DellaMorte:2010yp}. As we fix singlet quantum numbers for $p_y,p_z$ and ${\bf e}$, 
a stochastic treatment seems a priori justified, fluctuations are proportional to the exponentially 
suppressed higher momentum components, or to torelon contributions, which are expected to 
have higher energies on these volumes. The results in the next section have been obtained by using 
values of $n$ between 9 and 64, a choice justified a posteriori by the moderate statistical errors 
obtained. 
\begin{table}[!t]
\begin{center}
\begin{tabular}{|cccccc|}
\hline
Lattice&$L$&$T$&$N_\mathrm{conf}$&$N_\mathrm{lev}$&$d$\\[0.125cm]
\hline
${\rm A}_1$&   8  &  4  & 50  & 2 & 4  \\[0.125cm]
${\rm A}_2$&      &  5  & 50  & 2 & 5  \\[0.125cm]
${\rm A}_3$&      &  6  & 100 & 2 & 3  \\[0.125cm]
${\rm A}_4$&      &  8  & 100 & 2 & 4  \\[0.125cm]
${\rm A}_5$&      &  12 & 50  & 3 & $\left\{3,6\right\}$ \\[0.125cm]
\hline
${\rm B}_3$&   10 &  6  & 50  & 2 & 3  \\[0.125cm]
\hline
${\rm C}_1$&   14 &  10  & 100  & 2 & 5  \\[0.125cm]
\hline
\end{tabular}
\caption{Simulation parameters: $N_\mathrm{conf}$ is the number of configurations, 
$N_{\rm lev}$ is the number of levels and $d$ is the thickness of the thick time-slice 
used for the various levels.\protect\label{tab:par}}
\end{center}
\end{table}
\begin{table}[!t]
\begin{center}
\begin{tabular}{|cl|ccc|}
\hline
Lattice    & $n_1$ &  $Z^{({\bf p},+)}/Z$      &$Z^{({\bf p},+)}/Z^{({\bf 0},+)}$&
$a E_{\mathrm{eff}}^{({\bf {p}},+)}$\\[0.125cm]
\hline
${\rm A}_1$&  $1$  &$1.6(3)\cdot 10^{-3}$  &$1.04(21)\cdot 10^{-2}$&$1.14(5)$\\[0.125cm] 
${\rm A}_2$&  $1$  &$1.8(4)\cdot 10^{-3}$  &$2.0(5)\cdot 10^{-3}$  &$1.24(5)$\\[0.125cm]
${\rm A}_3$&  $1$  &$4.5(7)\cdot 10^{-4}$  &$4.7(7)\cdot 10^{-4}$  &$1.277(25)$\\[0.125cm]
${\rm A}_4$&  $1$  &$6.6(12)\cdot 10^{-5}$ &$6.6(12)\cdot 10^{-5}$ &$1.203(22)$\\[0.125cm]
${\rm A}_5$&  $1$  &$4.1(16)\cdot 10^{-7}$ &$4.3(17)\cdot 10^{-7}$ &$1.22(3)$\\[0.125cm]
\hline
${\rm B}_3$&  $1$  &$1.0(3)\cdot 10^{-3}$  &$1.0(3)\cdot 10^{-3}$  &$1.15(5)$\\[0.125cm]
           &  $2$  &$0.94(25)\cdot 10^{-4}$&$0.92(25)\cdot 10^{-4}$&$1.55(5)$\\[0.125cm]
\hline
${\rm C}_1$&  $1$  &$1.5(3)\cdot 10^{-4}$  &$1.5(4)\cdot 10^{-4}$  &$0.883(24)$\\[0.125cm]
           &  $2$  &$2.4(22)\cdot 10^{-5}$ &$2.4(22)\cdot 10^{-5}$& - \\[0.125cm]
\hline
\end{tabular}
\caption{Results for ratios of partition functions with momenta 
${\bf p} = [2\pi n_1/L, 0, 0]$. The effective energy $E_{\rm eff}^{({\bf p},+)}$ 
is defined as in Eq.~(\protect\ref{eq:Eeff}).\protect\label{tab:res}}
\end{center}
\end{table}
\vspace{-0.325cm}

\section{Results}
\vspace{-0.375cm}

We have simulated the SU(3) gauge theory discretized on the lattice by the Wilson action at 
$\beta=6/g^2_0=5.7$, and $\beta=5.85$  which correspond to a spacing of $0.17$~fm and $0.12$~fm 
respectively~\cite{Guagnelli:1998ud}. The spatial lengths are $1.4$ and $1.7$~fm, while 
time extends up to $2$~fm. The simulation parameters and the results are 
summarized in Tables~\ref{tab:par} and~\ref{tab:res}.
\begin{figure}[!t]
\begin{center}
\includegraphics[width=12.0cm]{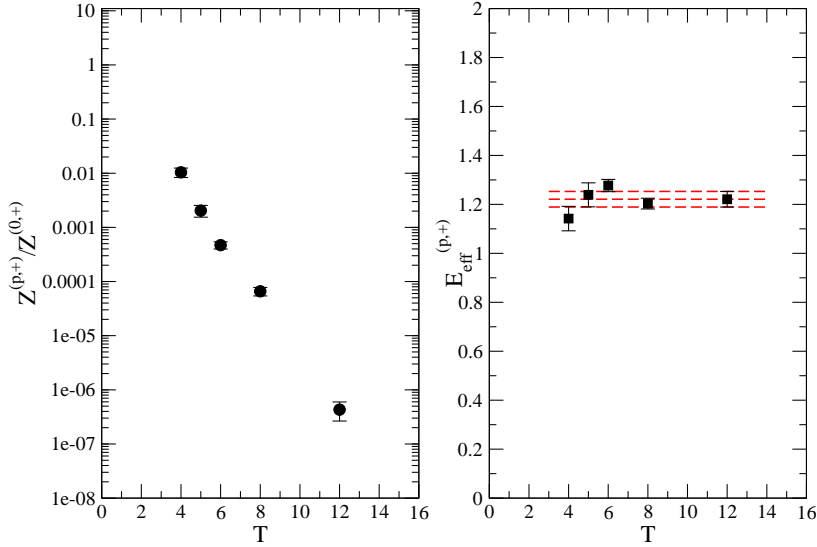}
\caption{
Left panel: ratio of partition functions $Z^{({\bf p},+)}/Z^{({\bf 0},+)}$ 
with momenta ${\bf p}=[2\pi/L,0,0]$ for the ${\rm A}$ lattices. Right panel: the 
corresponding effective energy as defined in Eq.~(\protect\ref{eq:Eeff}). The band is our 
best estimate, i.e. the one  extracted from the lattice with the longest 
time-extension.\protect\label{fig:fig0pp}}
\end{center}
\end{figure}

The primary quantity that we have computed is the ratio  $Z^{({\bf p},+)}/Z$ with 
$p_{2,3} = 0$, $p_1 = (2\pi/L) n_1$ and $n_1 = 1, 2$. The results for the A series 
are shown in the left plot of Figure~\ref{fig:fig0pp}. We fit them by 
assuming that a single state contributes, i.e. using an ansatz of the form 
$\ln(Z^{({\bf p},+)}(T)/Z(T))=A-BT$, such that $A$ yields the logarithm of the 
multiplicity of the state and $B$ its effective, finite momentum, 
energy  $E_{\rm eff}^{({\bf p},+)}$. At large time separations the lightest 
glueball with vacuum quantum numbers and momentum ${\bf p}$ is expected to 
dominate. The four points at the largest 
values of $T$ ($T/a=5,6,8,12$) in Figure~\ref{fig:fig0pp} are well described
by our ansatz, and the fit results for the multiplicity are well consistent with
a value of 1 excluding 2 and 3 by three and four standard deviations respectively.
We therefore define
\be
E_{\rm eff}^{({\bf p},+)}=-{{1}\over{T}} \ln \left[ {{Z^{({\bf p},+)}}\over{Z^{({\bf 0},+)}}} \right] \;,
\label{eq:Eeff}
\ee
for which the results are summarized in Table~\ref{tab:res} and shown on 
the right-hand plot of Figure~\ref{fig:fig0pp}. 

\begin{figure}[!t]
\begin{center}
\begin{tabular}{cc}
\includegraphics[width=7.0cm]{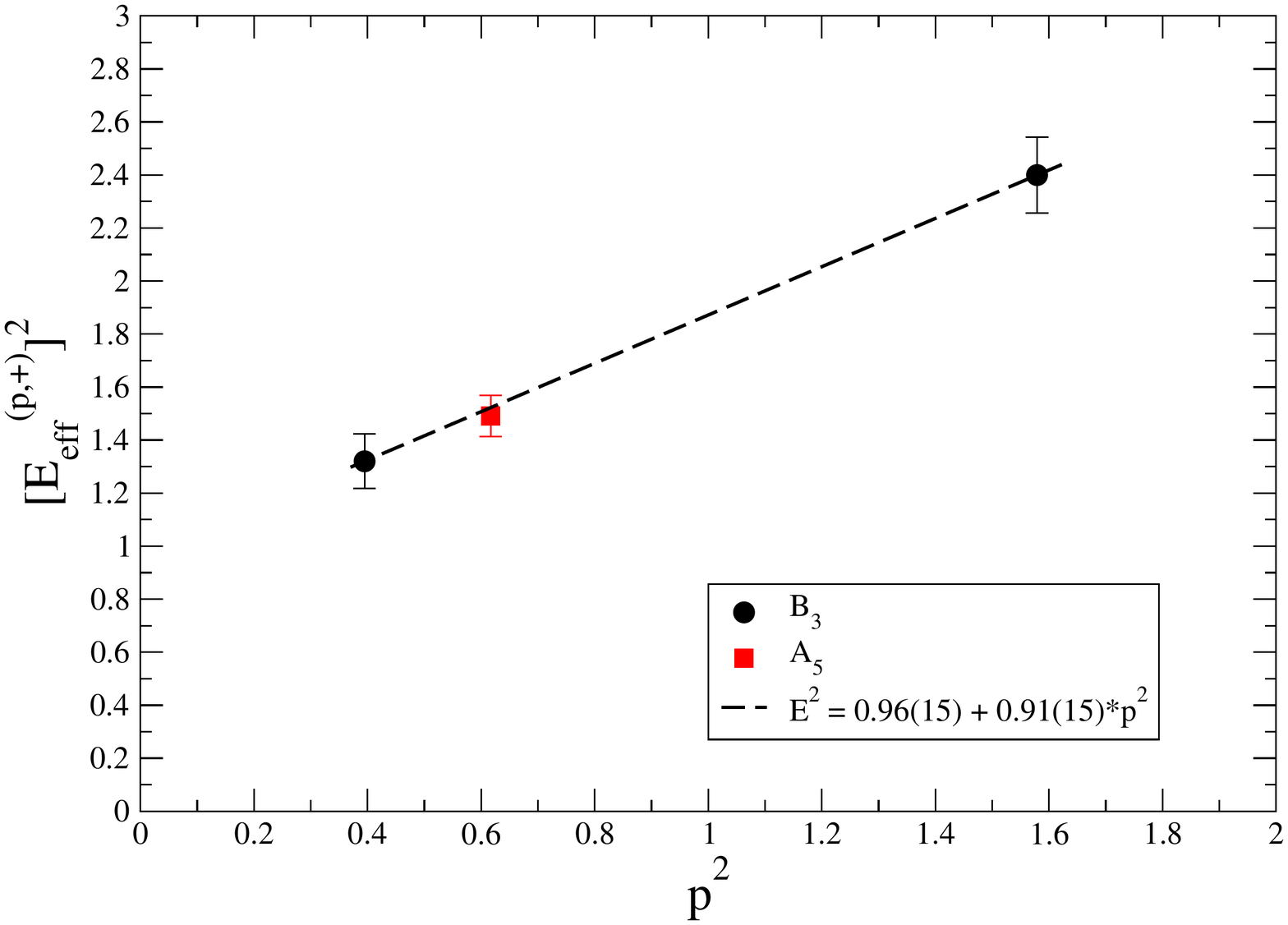} &
\includegraphics[width=7.0cm]{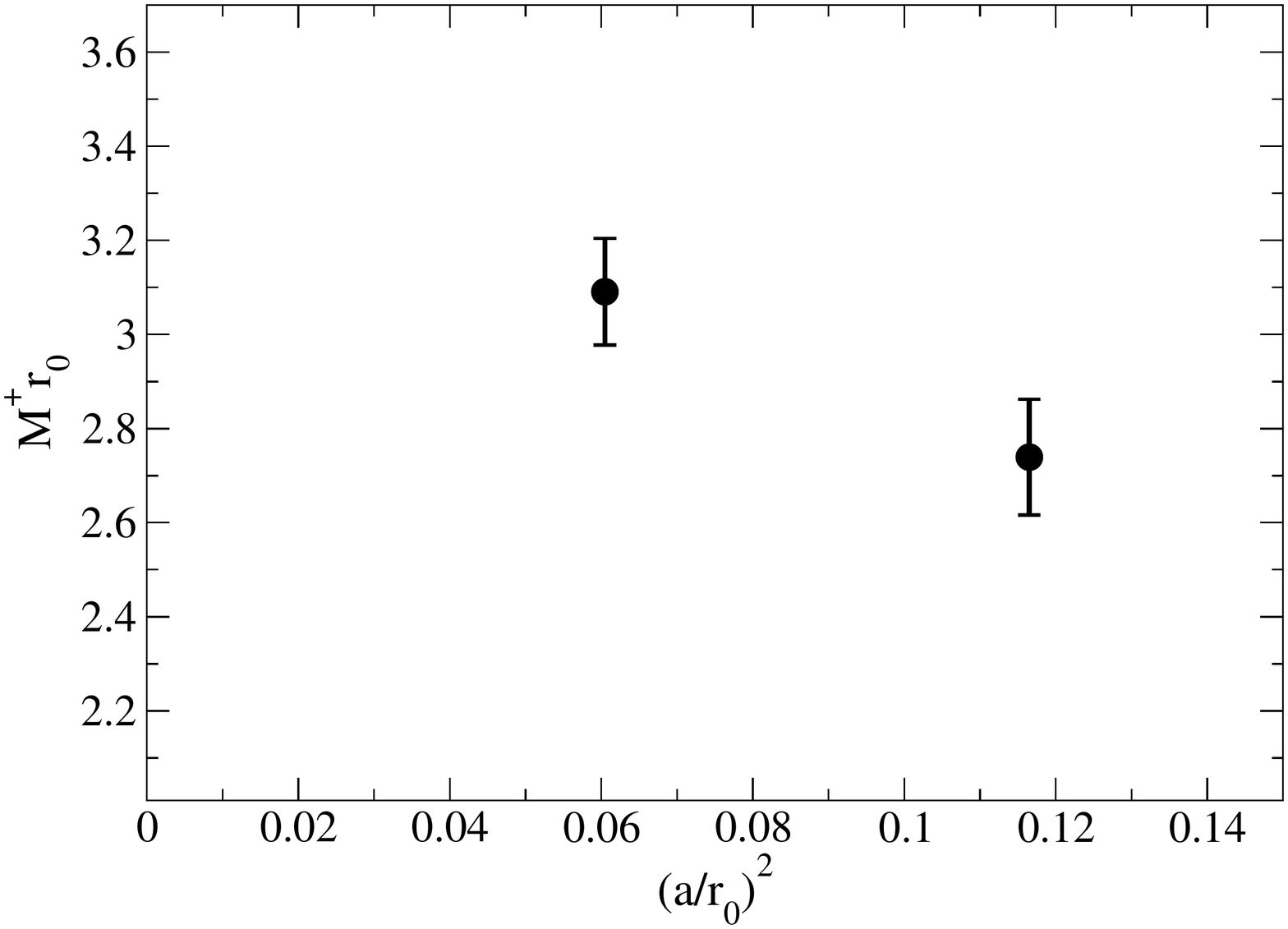}
\end{tabular}
\caption{Left panel: the effective energy squared from the $A_5$ (red square) 
and the $B_3$ (black circles) lattices. Right panel: results for $r_0 M^+$ at $a\simeq 0.17$ fm 
and $a\simeq 0.12$ fm.\protect\label{fig:cl}}
\end{center}
\end{figure}

The lattice ${\rm B}_3$ serves the purpose of assessing finite size effects. 
It has the same lattice spacing of the ${\rm A}$ series but a linear 
extension of $L=10$. The results for $n_1=1,2$ are reported in 
Table~\ref{tab:res}, and are plotted as a function of the momentum 
squared in the left plot of Figure~\ref{fig:cl}. The dashed line is a 
linear interpolation of the two black points  (circles) of the lattice 
${\rm B}_3$, while the red point (square) is our best result for the ${\rm A}$ series. 
It is rather clear that, within our statistical precision, 
finite volume effects are not visible in our data. It is also 
interesting to notice that, even if the values of the momenta are 
rather large, the continuum dispersion relation is well reproduced 
within our statistical errors.
We extract the mass of the lightest glueball from the expression
\be
M^+=\sqrt{(E_{\rm eff}^{({\bf p},+)})^2-{\bf p}^2} \;,
\ee
which, using the $T/a=12$ result and in units of the lattice spacing, gives
\be
M^+=0.935 \pm 0.042\;, \qquad[\beta=5.7]\;,
\ee
in good agreement with the estimate in Ref.~\cite{Vaccarino:1999ku} obtained at the same lattice spacing 
with the same discretization but within the standard approach. Finally, the lattice ${\rm C}_1$ is 
matched to ${\rm B}_3$ in volume but with a larger time extension (corresponding to $T/a=7$ at 
$\beta=5.7$) and more importantly  with a finer lattice resolution, namely $a=0.12$ fm. By making use 
again of Eq.~(\ref{eq:Eeff}) to extract $E_{\rm eff}^{({\bf p},+)}$, we get from the continuum dispersion relation
\be
M^+=0.760 \pm 0.028\;, \qquad[\beta=5.85]\;, 
\ee
in units of the lattice spacing. The results, measured in units of the 
scale $r_0$~\cite{Guagnelli:1998ud}, are 
collected in Fig.~\ref{fig:cl}, where they are plotted as a function of $(a/r_0)^2$ as the 
leading discretization effects should be quadratic in the lattice spacing. 
\vspace{-0.325cm}

\section{Conclusions}
\vspace{-0.375cm}

We have discussed how
the relative contributions to the partition function, due to states carrying a given set
of quantum numbers associated with the exact symmetries of a field theory, can be
expressed by ratios of path integrals with different boundary conditions in the time
direction. From an algorithmic point of view, the composition properties
of the projectors can be exploited to implement a hierarchical multi-level integration
procedure which solves the problem of the exponential (in time) degradation of the
signal-to-noise ratio.
                          
Within this approach we have performed a precise lattice computation
of  the mass of the lightest glueball with vacuum quantum numbers in the SU(3) 
Yang-Mills theory at two values of the lattice spacing corresponding to 0.17 and 0.12 fm.
The algorithm works as expected, and we have been able to follow the exponential
decay up to separations of 2 fm, while keeping the error on the effective mass approximatively 
constant as a function of time. 

Cutoff effects appear to be rather large and tend to significantly decrease the estimate of the
glueball mass at finite lattice spacing, as expected and also observed in previous lattice 
computations~\cite{Hasenbusch:2004yq,Necco:2003vh}.
We are now in the process of repeating the calculation presented at a finer lattice
resolution of 0.1 fm, which should eventually allow us to properly assess the magnitude of 
discretization effects.\\[-0.375cm]

The simulations were performed at CILEA, at the Swiss National 
Supercomputing Centre (CSCS) and at the J\"ulich Supercomputing Centre (JSC). 
We thankfully acknowledge the computer resources and technical support provided 
by all these institutions and their technical staff.
\vspace{-0.375cm}

\end{document}